\documentclass[pre,aps,twocolumn,showpacs,floatfix,superscriptaddress,nolongbibliography]{revtex4-2}

\usepackage{graphicx}
\usepackage{bm}
\usepackage{color}
\usepackage{amssymb,amsmath}
\usepackage[normalem]{ulem}
\usepackage{ragged2e}
\usepackage{adjustbox}
\usepackage{multirow}
\usepackage{booktabs}
\usepackage[table]{xcolor}
\usepackage{tcolorbox}
\usepackage{tabularx}
\usepackage{array}
\usepackage{colortbl}
\usepackage{hyperref}
\hypersetup{colorlinks=true, citecolor=blue, urlcolor=blue, linkcolor=blue}
\tcbuselibrary{skins}
\definecolor{Salmon}{RGB}{250,128,114}

\newcolumntype{Y}{>{\raggedleft\arraybackslash}X}

\tcbset{tab1/.style={\centering, fonttitle=\bfseries\small,fontupper=\normalsize\sffamily,
		colback=yellow!10!white,colframe=red!75!black,colbacktitle=Salmon!40!white,
		coltitle=black,center title,freelance,frame code={
			\foreach \n in {north east,north west,south east,south west}
			{\path [fill=red!75!black] (interior.\n) circle (3mm); };},}}

\tcbset{tab2/.style={enhanced,fonttitle=\bfseries,fontupper=\small\sffamily,
		colback=yellow!10!white,colframe=red!50!black,colbacktitle=Salmon!40!white,
		coltitle=black,center title}}

\begin{document}

\title{\Large Controlling complex rhythms: A hierarchical approach to limit cycle switching}

\author{Sandip Saha}
\email{sandips@ncbs.res.in; sahasandip.loknath@gmail.com}
\thanks{These authors contributed equally to this work.} \thanks{Corresponding author}
\affiliation{National Centre for Biological Sciences -- Tata Institute of Fundamental Research (NCBS-TIFR), Bangalore, India}
\affiliation{Physics and Applied Mathematics Unit, Indian Statistical Institute, 203 B. T. Road, Kolkata 700108, India}

\author{Suvam Pal}
\email{suvamjoy256@gmail.com}
\thanks{These authors contributed equally to this work.}
\affiliation{Physics and Applied Mathematics Unit, Indian Statistical Institute, 203 B. T. Road, Kolkata 700108, India}

\author{Dibakar Ghosh}
\email{dibakar@isical.ac.in}
\affiliation{Physics and Applied Mathematics Unit, Indian Statistical Institute, 203 B. T. Road, Kolkata 700108, India}

	

\begin{abstract}
    \par Limit cycles are self-sustained, closed trajectories in phase space representing (un)-stable, periodic behavior in nonlinear dynamical systems. They underpin diverse natural phenomena, from neuronal firing patterns to engineering oscillations. The presence of multiple concentric limit cycles reflects distinct behavioral symmetries within a system. In this work, we investigate the hierarchical dynamical transitions from one limit cycle to another, driven by oscillatory excitation while preserving other system properties. We demonstrate that controlling multirhythmicity through hierarchical, stepwise periodic modulation enables reliable switching between rhythmic states. This hierarchical control framework is crucial for applications in neuro-engineering and synthetic biology, where precise, robust modulation of complex rhythmic behaviors enhances system functionality and adaptability. 
\end{abstract}
	

\maketitle

\section{Introduction}
In recent decades, the emergence of interdisciplinary fields has underscored the necessity of constructing prototype mathematical models to gain a generalized understanding of complex natural phenomena and real-world challenges~\cite{slack199, rand2021geometry, bergen2020generalizing}. One common approach involves starting with a known simple model that captures the essential features of a system~\cite{rand2021geometry, goldbook}. Although such models are idealized, they often provide valuable insights and serve as a basis for understanding more complex behaviors~\cite{rand2021geometry, goldbook}. These simple models, originally developed within specific disciplines--for instance, mechanical systems in physics~\cite{kbbook, Esmailzadeh1997, landau_book, Eichler2011}--have found broad applicability across scientific domains~\cite{kbbook, landau_book, goldbook, strogatz, epstein, Jenkins-2013-PhysicsReports}. For example, models originally used to describe mechanical vibrations~\cite{kbbook, Faraday_1831_paper} may be adapted to study chemical oscillations in far-from-equilibrium systems~\cite{epstein} or biological rhythms~\cite{goldbook} in living organisms~\cite{goldbook, menakar, jewett, cirgoldbeter1995, pal2024directional, ghosh2024amplitude}.

A typical strategy in this context involves formulating a mathematical model of the system under investigation~\cite{rand2021geometry, raju2024geometrical}, regardless of the field, and comparing it with a known simple model under a restricted set of parameters~\cite{rand2021geometry, raju2024geometrical}. By gradually relaxing these constraints, one can observe how the system reacts as it departs from the idealized behavior~\cite{rand2021geometry, kbbook}. This allows researchers to probe the dynamics near the well-understood regime of the simple model. This methodology aligns with a fundamental principle in science: real-world systems are inherently imperfect~\cite{waddington2014strategy} and rarely match ideal models exactly~\cite{kaneko2006life,waddington2014strategy}. Therefore, using a known simple model as an approximation offers a useful starting point~\cite{rand2021geometry, raju2024geometrical}. Looking into the deviations from the ideal situation opens a new avenue to probe the critical and technical aspects of the real-life systems~\cite{rand2021geometry, raju2024geometrical}. Also, from this we get deeper and systematic insights regarding this model.

From a mathematical standpoint, a well-known example illustrating the above approach is the Krylov--Bogoliubov method~\cite{kbbook, len0, strogatz}, which is widely used in the analysis of nonlinear systems~\cite{kbbook, len0}. The core idea behind the method is conceptually straightforward and forms part of standard textbook material~\cite{kbbook, len0, strogatz}. Regardless of whether the nonlinearity in a system is weak or strong, the method begins by approximating the system near a simple, well-understood model--typically the harmonic oscillator. Once this baseline is established, a nonlinearity control parameter can be gradually introduced to investigate how the system's behavior--such as amplitude and phase--deforms in the presence of nonlinear effects. In the case of strongly nonlinear systems, a common technique involves introducing a small parameter (often denoted by $\epsilon$) in front of the nonlinear terms. This allows the system to be treated perturbatively, and the parameter can ultimately be set to unity after the analysis is complete. This strategy enables a controlled exploration of nonlinear dynamics while leveraging the simplicity of the base model~\cite{len0, strogatz}.

In a similar manner, to understand complex phenomena across various scientific domains~\cite{Goldbeter84, Decroly1982,bergen2020generalizing,waddington2014strategy}, it is often essential to begin with simplified models~\cite{gly1,lavrova2009,bergen2020generalizing,raju2024geometrical}. This approach allows researchers to develop a robust theoretical framework by systematically exploring the behavior of a system through a limited and well-defined parameter space~\cite{rand2021geometry, raju2024geometrical}. In biology, for instance, simplified kinetic models, such as the Sel’kov model~\cite{len0}, is used to study the intricate dynamics of glycolysis~\cite{Goldbeter84,Decroly1982}. Similarly, in chemistry, models of oscillatory reactions help explain far-from-equilibrium processes~\cite{beato2007,zhabotinsky1964}, such as color-changing reactions arising from nonlinear chemical dynamics~\cite{sen2008,beato2007,zhabotinsky1964}. In physics, prototype models like the complex Ginzburg–Landau~\cite{aranson2002world} equation serve as foundational tools to analyze large-scale synchronization and interaction phenomena~\cite{arenas2008synchronization}. Attempting to understand such complex systems by directly employing arbitrary models or exploring vast parameter spaces without structure can lead to misleading conclusions~\cite{arenas2008synchronization}. Hence, a step-by-step modeling strategy is necessary to ensure scientific rigor and meaningful insight.

Understanding complex systems often requires the use of simplified or prototype model systems, which allow for comparative insights through the lens of dynamical systems theory~\cite{strogatz,rand2021geometry}. One central approach in this field involves analyzing the stability of fixed points to characterize system dynamics~\cite{strogatz}. In certain cases, complexity emerges in the form of multiple fixed points/oscillations--referred to as multistability~\cite{Pisarchik2014}--or from a single unstable fixed point surrounded by one or more stable limit cycles, a phenomenon known as multirhythmicity~\cite{saha2020systematic}. This latter behavior is particularly intriguing, and challenging, as it involves concentric limit cycles divided by an intermediate unstable limit cycle~\cite{saha2020systematic}. The unstable cycle effectively separates the basins of attraction of the stable ones, adhering to Poincar\'e index theory, which mandates that the total index for a bounded system sums to unity\cite{smith1984poincare}.

The application of control schemes to achieve a desired amplitude of oscillation in multirhythmic systems remains in limelight, even since the initial discovery of multirhythmic behavior\cite{Goldbeter84,Goswami2008,Pisarchik2014}. Despite various control schemes--such as self-feedback~\cite{biswas_chaos_2017}, conjugate feedback\cite{biswas_pre_2016}, delayed feedback~\cite{k-dsr,cheage2012dynamics}, low-pass filtering~\cite{biswas_pre_2019} and parametrically modulated nonlinearity~\cite{saha2021suppressing}--proposed in the literature to address multirhythmicity, these methods are often limited by their dependence on specific systems and parameter regimes. A universally robust control mechanism remains elusive. Among the schemes discussed above, parametric modulation~\cite{saha2021suppressing} relies solely on the system's intrinsic frequency to achieve control. This approach exploits the resonance arising from the interplay between the system's inherent rhythmic dynamics and the modulation of its parameters, without introducing any additional dimensions to the system. To this end, we pose a fundamental question: {\it Is the control process a single-step transition that enables the direct suppression of higher-order rhythmic variants into a monorhythmic state--the simplest form often sought for studying biophysical processes—or does it require a hierarchical, step-by-step approach to achieve monorhythmicity?} A recent study by Guo et al.~\cite{guo2024emergent} explored a similar class of systems but did not address this foundational question. In this work, we aim to investigate and provide insight into this issue by analyzing a prototypical model that captures higher-order rhythmic dynamics. This model may offer broader experimental relevance in the future. Crucially, we argue that the control of high-order rhythmic behavior in complex systems is unlikely to be achieved through a single-step intervention. Rather, a hierarchical and sequential approach is necessary--a principle that forms the basis of the methodology developed and presented in the subsequent sections.

In what follows, Sec.\ref{sec1} discusses why the delayed feedback mechanism is not a suitable candidate for control schemes, as it induces multirhythmicity in an otherwise monorhythmic system. In Sec.\ref{sec2}, we adopt a more efficient approach that leverages the system’s intrinsic parameters to investigate the hierarchical process of controlling rhythmicity, using a recently proposed prototype multirhythmic model. Building upon this, Sec.\ref{eff-pot-sec} explores the complexity and underlying dynamics of multirhythmic systems through the framework of an effective potential. In Sec.\ref{stability}, we analyze the switching phenomena between stable limit cycles under internal periodic modulation and examine their stability on the van der Pol plane. This analysis supports the hierarchical control mechanism via a mathematical reconstruction based on a simple transformation. Additionally, Sec.\ref{basin} discusses how the domain of initial conditions varies with modulation strength. Finally, Sec.\ref{conclusion} summarizes the hierarchical control framework and presents our concluding remarks.

\section{Delay induced multirhythmicity}\label{sec1}
\begin{figure}
    \centering
    \includegraphics[width=\linewidth]{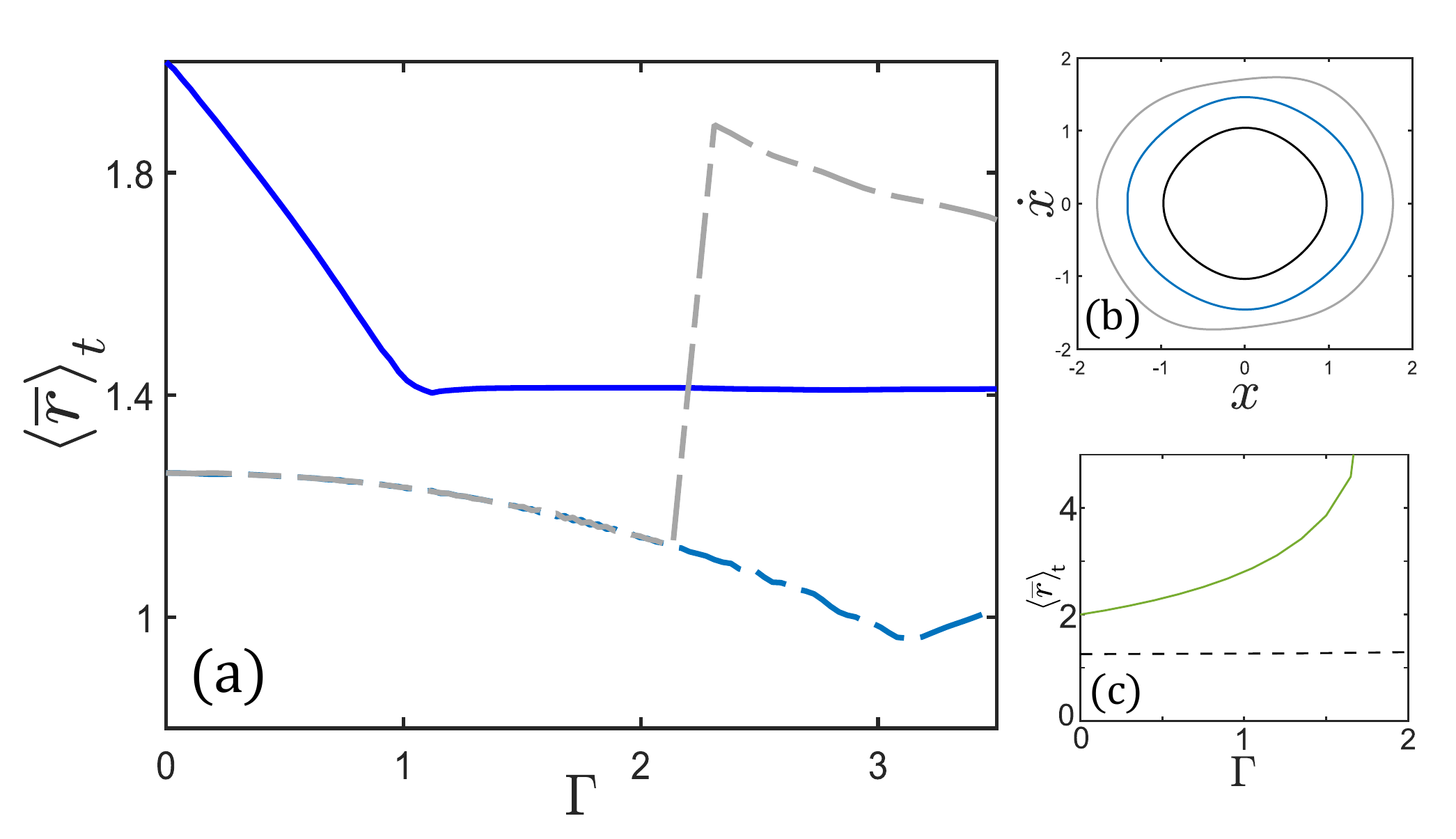}
    \caption{Behavior of the amplitude ($\langle \overline{r} \rangle_t$) in terms of the strength of the periodic modulation ($\Gamma$) for the dynamics in Eq.~\eqref{shm-non1-delay} with and without delay term. In (a), we describe the behavior of \textbf{$\langle \overline{r} \rangle_t$} in terms of $\Gamma$ for $\Omega=2$, where solid-blue line depicts without delay scenario. Whereas, dashed lines show the behavior with delay term. In (b), we represent the limit cycles on $x-\dot{x}$-plane. Here, solid blue line describes the dynamics without delay; on the other-hand solid black and gray lines describe the with delay term for two different sets of initial conditions $(x_0,v_0)=(1.056,-0.8576)~\text{and}~(1.576,1.037)$ respectively, for $\Gamma=3.3$. Finally in (c), we show $\Gamma$ vs. $\langle \overline{r} \rangle_t$ for $\Omega=4$, where solid-line signifies the dynamics without delay term, however dashed-line is for dynamics with delay term, indicating the suppression of uncontrolled growth in amplitude. For these numerical estimations we set $\mu=0.1,~\tau=0.623$ and $K=0.1$.}
    \label{fig1}
\end{figure}

Starting with the minimal model of a simple van del Pol (vdP) oscillator, driven by parametrically weak non-linear damping term~\cite{penvo,saha2021suppressing} including self-feedback term incorporating time delay~\cite{k-dsr}, that reads
\begin{equation}\label{shm-non1-delay}
    \Ddot{x}+\mu[1+\Gamma \cos(\Omega t)](-1+x^2)\dot{x}+x-Kx(t-\tau)=0,
\end{equation}
where $\mu$ is the non-linear damping strength and $\Gamma$ describes the amplitude of the periodic modulation~\cite{penvo,saha2021suppressing}. Note that, $\Gamma=0,~K=0$ recovers the dynamics of classical vdP oscillator, which exhibits a stable limit cycle in the $x-\dot{x}$ phase plane. This is evident from Figs.~\ref{fig1}(a) and \ref{fig1}(c), in $\Gamma\rightarrow 0$ limit, dynamics emerges limit cycle with amplitude $\langle \overline{r} \rangle_t\approx 2$. For instance, we evaluate time-averaged amplitude, \textit{i.e.}, $\langle \overline{r} \rangle_t$ after a large transient time.

In the absence of the delay term, the dynamics in Eq.~\eqref{shm-non1-delay} emerges two concentric limit cycles branches for modulation frequencies $\Omega=2$ and $\Omega=4$, which is in stark contrast to the single limit cycle behavior of the conventional vdP oscillator. Such branching due to parametric modulation is illustrated in Fig.~\ref{fig1}. Specifically, Fig.~\ref{fig1}(a) shows the dependence of amplitude $\langle \overline{r} \rangle_t$ on the modulation strength $\Gamma$ for $\Omega=2$. Here solid blue curve indicates that $\langle \overline{r} \rangle_t$ initially decreases and then saturates to a nearly constant value as $\Gamma$ increases, indicating suppression of oscillation amplitude. In contrast to this, Fig.~\ref{fig1}(c) displays a different scenario for $\Omega=4$, where the solid green curve describes a non-linear growth of the average amplitude with increasing modulation strength, signifying enhanced resonance effect.

The presence time delay ($K,\tau \neq 0$) plays a key role in controlling the abrupt growth in amplitude for $\Omega=4$ (see Fig.~\ref{fig1}(c), in contrast to $\Omega=2$. The delay acts to regulate and constrain the abrupt increment in oscillation amplitude that occurs for larger modulation strengths. However, interplay between the delay and dynamics of lower frequency leads the limit cycle to exhibit further branching, that is evident from Fig.~\ref{fig1}(a) (see dashed curve). Also in Fig.~\ref{fig1}(b), we demonstrate the dynamics with and without delay term, where solid blue line corresponds to the limit cycle without delay term; whereas, solid black and gray lines correspond to the smaller and larger limit cycle. However, it should be remembered that incorporation of time-delay introduces additional degrees of freedom, rendering the system to be infinite-dimensional. Consequently, while delay can help regulate multi-stable limit cycles~\cite{k-dsr,cheage2012dynamics}, it also imposes complexity that may limit its consistency and practical utility as a control mechanism. In contrast to this, we provide a systematic investigation of controlling rhythmicity for a prototype multirhythmic model in the next section.

\section{Multirhythmicity : prototype model and controlling mechanism}\label{sec2}
Let us consider a multi-rhythmic variant of vdP oscillator with extended non-linear damping terms~\cite{saha2020systematic, guo2024emergent}, which reads
\begin{align}\label{new-model}
    \Ddot{x}+&\mu[1+\Gamma \cos(\Omega t)]\times\nonumber\\
    &(-1+x^2-\alpha x^4+\beta x^6-\gamma x^8+\delta x^{10})\dot{x}+x=0,
\end{align}
where multiplicative constants $\alpha, \beta, \gamma, \delta \in \mathcal{R}^+$. Additionally, the system exhibits three limit cycles with different amplitudes with $\Gamma=0$. Furthermore, an extensive study of the dynamics in Eq.~\eqref{new-model} reveals three concentric limit cycles in the absence of periodic modulation, which will be discussed in the next subsection.

Moreover, the presence of the periodic nonlinear damping term causes the system to exhibit six resonating cycles for $\Omega=(2,4,6,8,10,12)$ for the three stable limit cycles as it is evident from the classical vdP oscillator~\cite{penvo,saha2021suppressing} in the previous section. These modes of resonances are coming directly from the singularity presence in the amplitude-phase dynamics (Eqs.~\ref{dyn-avg-amp} and \ref{dyn-avg-phi}). To find the amplitude-phase dynamics, we use the slower lime scale present in the system due to weak nonlinearity, \textit{i.e.}, $\mathcal{T}\sim \mathcal{O}(\mu)$. Essentially, this timescale $\mathcal{T}$ allows us to adopt the perturbative approaches such as Krylov-Bogoliubov method~\cite{kbbook,len0,strogatz} to address the limit cycles with oscillating amplitudes in the polar plane. Starting with the ansatz $x(t)=r(t)\cos(t+\phi(t))$ and $\dot{x}(t)=-r(t)\sin(t+\phi(t))$, where $(r,\phi)=(\sqrt{x^2+\dot{x}^2},-t+\tan^{-1}(-\dot{x}/x))$. Compared to the underlying dynamics on the $x-\dot{x}$ phase-space, $r$ and $\phi$ are slowly varying functions of time, since $0<\mu \ll 1$. Furthermore, $r(t)~\text{and}~\phi(t)$ can be considered as $r(t)\approx \overline{r}(t)+\mathcal{O}(\mu)$ and $\phi(t)\approx \overline{\phi}(t)+\mathcal{O}(\mu)$. Here, we use the definition that average of a function, $\mathcal{H}(x,\dot{x})$ (say), over a period $2\pi$ is conventionally denotes as $\overline{\mathcal{H}}(t)=\dfrac{1}{2\pi}\int_0^{2\pi}\mathcal{H}(s)~ds$. Finally, we obtain the following dynamics of average amplitude $\overline{r}(t)$ and the average phase $\overline{\phi}(t)$ as,
\begin{align}
    \dot{\overline{r}}(t)=&\dfrac{\mu}{1024}\overline{r}(512-128\overline{r}^2+64\alpha \overline{r}^4\nonumber\\
    &-40\beta \overline{r}^6+28\gamma \overline{r}^8-21\delta \overline{r}^{10})+\left.\mathcal{A}\right|_{\Omega}(\overline{r},\overline{\phi},\Gamma),\label{dyn-avg-amp}\\
    \dot{\overline{\phi}}(t)=&\left.\mathcal{B}\right|_{\Omega}(\overline{r},\overline{\phi},\Gamma).\label{dyn-avg-phi}
\end{align}
We provide the functional forms of $\left.\mathcal{A}\right|_{\Omega}(\overline{r},\overline{\phi},\Gamma)$ and $\left.\mathcal{B}\right|_{\Omega}(\overline{r},\overline{\phi},\Gamma)$ in Appendix.~\ref{appendixA}. 

\section{Description of complexity through effective potential}\label{eff-pot-sec}
\begin{figure}
    \centering
    \includegraphics[width=0.8\linewidth]{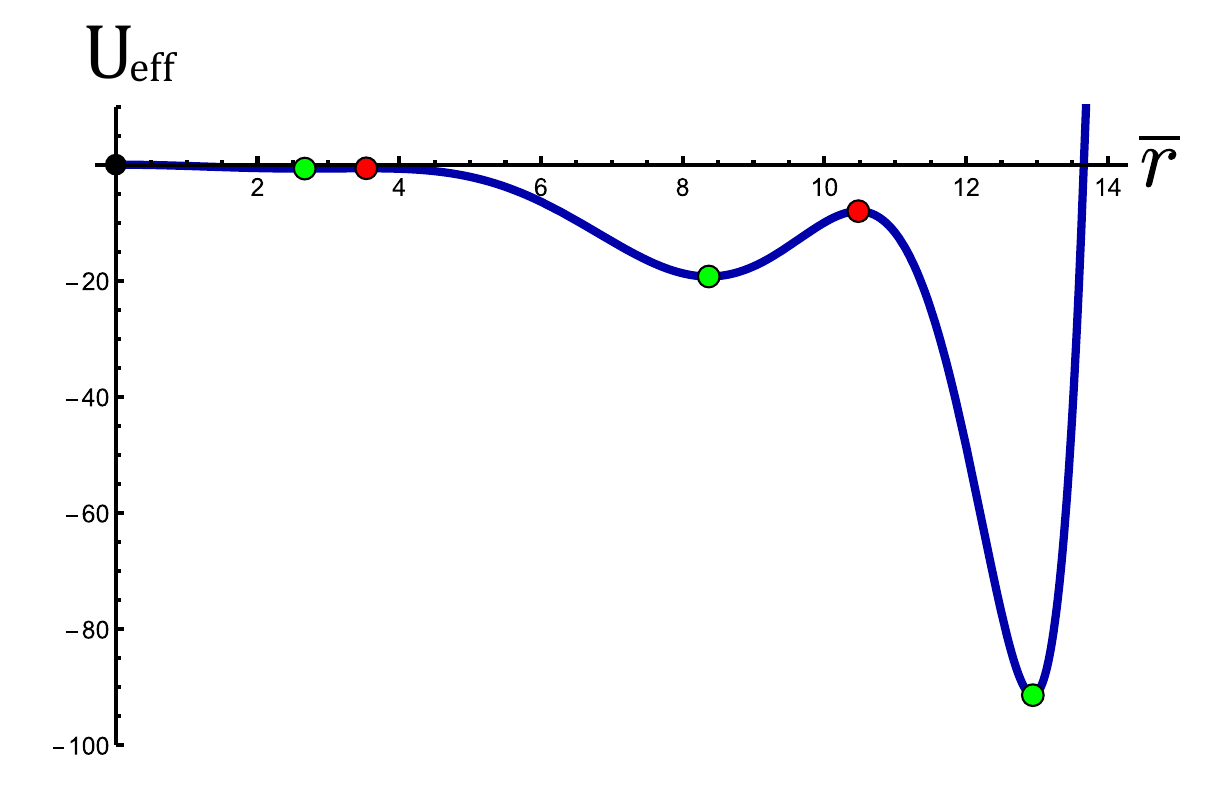}
    \caption{A schematic representation of the effective potential $U_{eff}$ (see Eq.~\ref{eff-pot}) with $\alpha=0.144,~\beta=5\times 10^{-3},~\gamma=5.862\times 10^{-5}$ and $\delta=2.13\times 10^{-7}$. The minima (stable limit cycles) of the potential are indicated with small circles (colored in green), whereas local maxima (unstable limit cycles) are indicated with small circles (colored in red).}
    \label{fig2}
\end{figure}

By setting $\Gamma=0$, additive terms in Eqs.~\eqref{dyn-avg-amp} and \eqref{dyn-avg-phi}, \textit{i.e.}, $\mathcal{A},~\mathcal{B}$ become zero; so that one can write the dynamics of average amplitude (see Eq.~\ref{dyn-avg-amp}) in terms of the effective potential $U_{eff}$. Due to the gradient nature of the effective potential, Eq.~\eqref{dyn-avg-amp} can be written in the following way
\begin{equation}
    \dot{\overline{r}}(t)=-\dfrac{\partial}{\partial \overline{r}} U_{eff}(\overline{r}),
\end{equation}
with 
\begin{align}\label{eff-pot}
    U_{eff}(\overline{r})=&-\dfrac{1}{4}\overline{r}^2+\dfrac{1}{32}\overline{r}^4-\dfrac{\alpha}{96}\overline{r}^6\nonumber\\
    &+\dfrac{5\beta}{1024}\overline{r}^8-\dfrac{7\gamma}{2560}\overline{r}^{10}+\dfrac{7\delta}{4096}\overline{r}^{12}.
\end{align}
Note that, we scale $U_{eff}$ with $\mu$ (strength of the non-linear drift). Evidently from Fig.~\ref{fig2}, the dynamics exhibit five concentric limit cycles surrounding an unstable focus at the origin (indicated with a small circle in black). Moreover, stability of the limit cycles can be understood in the perspective of the effective potential: (i) three of them are stable (indicated in green color in Fig.~\ref{fig2}), (ii) the other two are unstable (colored in red). Intuitively, the unstable limit cycles serve as a basin boundary of the stable limit cycles. 

In line with our objective, the key question is whether periodic modulation can consolidate all existing limit cycles into a single, unique limit cycle, or whether a systematic, stepwise approach is required to converge toward a unique limit cycle. To address such a scenario, we analyze the correlation between $\Gamma$ and the mitigation of limit cycles in the next section.


\section{Trade off between $\Gamma$ and resonating cycles : facets of stability}\label{stability}
\begin{figure*}[t]
    \centering
    \includegraphics[width=\linewidth]{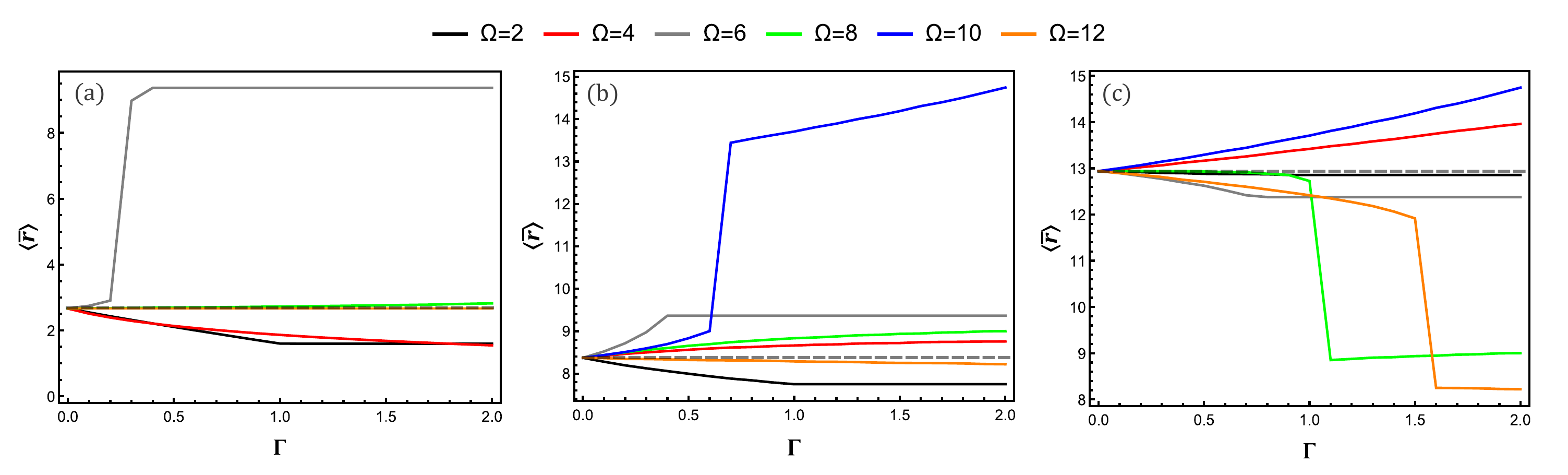}
    \caption{We illustrate the variation of the amplitude for three different sets of initial conditions, \textit{i.e.}, initial conditions have been taken from the domain around the potential depths (cf. Fig.~\ref{fig2}). Here, we observe that the sharp dependence on initial conditions drives the limit cycles for different modulation frequencies. For instance, in (a), we observe that the periodic modulation drives the limit cycle (for $\Omega=6$) into the adjacent stable domain, while other limit cycles remain stabilized near the domain corresponding to their initial conditions. In contrast to this, we do not observe any switching in amplitude to the consecutive stable domains for this particular frequency, as described in (b) and (c). Essentially, this point to the controlling feature for $\Omega=6$. A similar behavior is observed for $\Omega = 10$, where initial conditions near the intermediate-amplitude limit cycle (see panel b) evolve toward the large-amplitude cycle. In contrast, for $\Omega = 8$ and $12$, the dynamics exhibit amplitude suppression, with trajectories initialized in the large-amplitude domain eventually saturating at the intermediate limit cycle.}
    \label{fig3}
\end{figure*}
As discussed in the previous section, in the absence of periodic modulation, the system supports three distinct limit cycles, contingent on the basin of initial conditions. The introduction of periodic forcing induces amplitude modulation and inter-cycle switching~\cite{saha2021suppressing}, as clearly illustrated in Fig.~\ref{fig3}. The amplitude response exhibits strong dependence on the initial state, leading to mode-specific variations. Notably, in Fig.\ref{fig3}(a), a pronounced amplitude discontinuity is observed for $\Omega = 6$ around $\Gamma = 0.15$, where the system abruptly transitions to an intermediate-amplitude branch and remains stabilized thereafter. We also observe that $\Omega=6,8$ exhibit resonance for that particular domain of initial condition, whereas other frequencies show anti-resonance. However, with initial conditions from the intermediate regime, $\Omega=4,6,8,10$ exhibit resonance and others show anti-resonance (see Fig.~\ref{fig3}(b)). Amplitude of $\Omega=10$, shows an abrupt jump after $\Gamma=0.6$ and settles to the furthest amplitude. A similar amplitude switching is observed in Fig.~\ref{fig3}(c) for $\Omega = 8$ and $\Omega = 12$, where the dynamics originate near the outermost amplitude branch. 

The most important part of this report through the lens of observation, is that, Fig.~\ref{fig3} reveals the control in the amplitude of oscillation. For instance, consider the limit cycles regarding $\Omega=6$, in Fig.~\ref{fig3}(a), we observe that the amplitude of oscillation settles down in the intermediate domain of the potential structure. However, for the initial condition in the vicinity of the second depth, the amplitude becomes stable in that domain (see Fig.~\ref{fig3}(b)). Similarly, if the initial conditions lie near the third well of the potential, the oscillation remains confined there. This indicates that for $\Omega=6$, the switching of stability of the amplitude is strongly correlated with the domains of initial conditions. In essence, the strength of the periodic modulation becomes a cornerstone to shape the stability landscape, and that determines where the oscillation stabilizes for each value of $\Omega$, provide us a stepwise controlling scheme of multirhythmicity. Furthermore, this allows us to investigate the correlation between $\Gamma$ and initial conditions.

\subsection{Basin of attraction}\label{basin}
\begin{figure}[t]
    \centering
    \includegraphics[width=\linewidth]{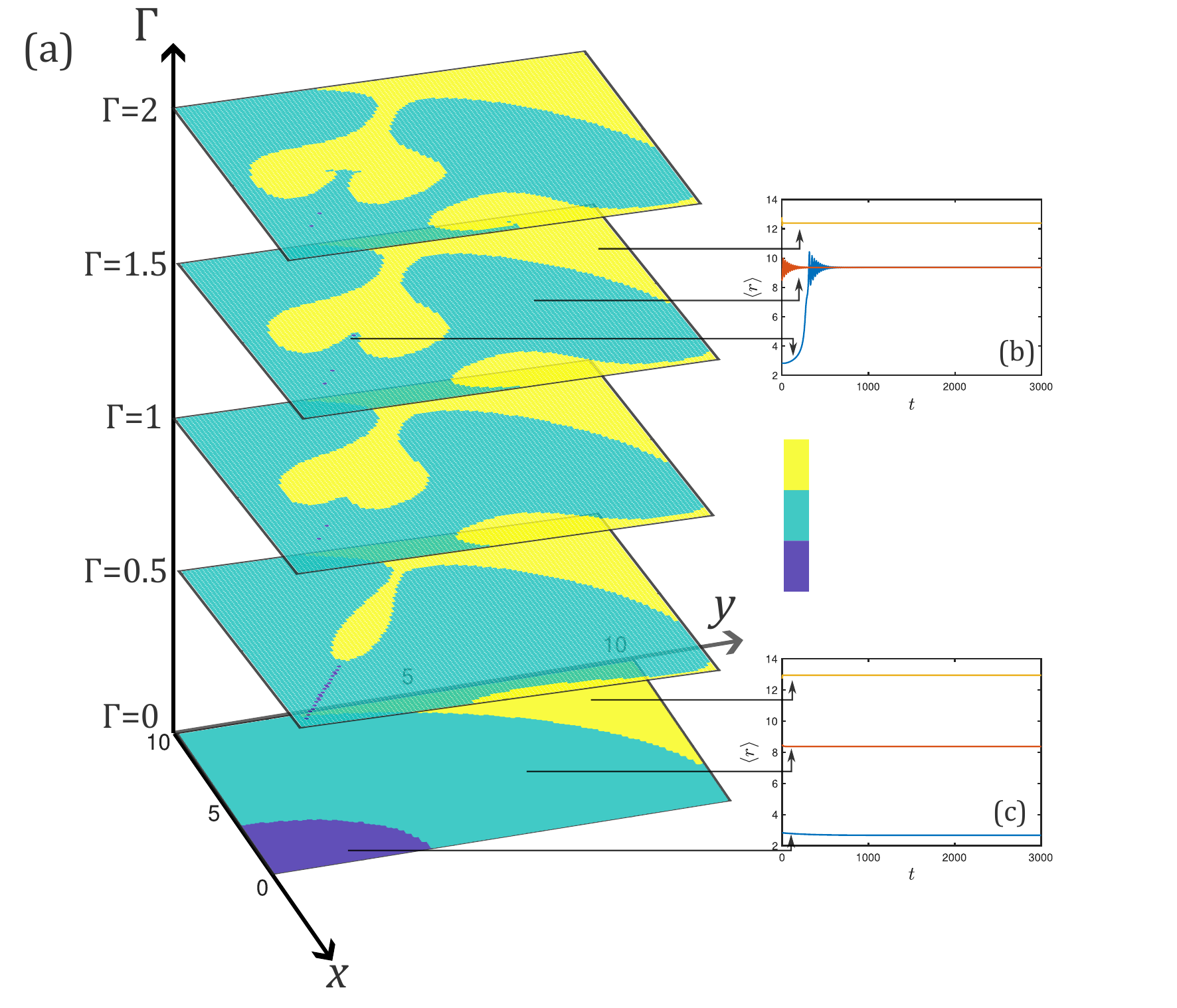}
    \caption{We investigate how the amplitude of the limit cycle at $\Omega = 6$ evolves as a function of the modulation strength $\Gamma$, focusing on its dependence on the initial conditions. In the absence of modulation ($\Gamma = 0$), the system exhibits multistability, characterized by the presence of three distinct stable limit cycles. These correspond to three separate domains in the space of initial conditions, which can be identified by examining the underlying effective potential landscape (refer to Fig.~\ref{fig2}). To visualize this, we categorize the initial conditions based on the amplitude of the resulting limit cycles. In Fig.~\ref{fig2}, the blue region indicates initial conditions that lead to small-amplitude oscillations, while the yellow region corresponds to those resulting in large-amplitude cycles. The region lying between these two extremes—represented by an intermediate color—captures initial conditions that give rise to limit cycles with moderate amplitudes. This clear separation of dynamic outcomes emphasizes the system’s inherent multistability at $\Gamma = 0$. As the modulation strength $\Gamma$ increases (panels b and c), we observe significant deformation and reorganization of these domains. The boundaries between the different regions shift, and the overall distribution of initial conditions evolves, indicating a growing level of control exerted by the periodic modulation. This progressive transformation illustrates how modulation not only influences the stability of limit cycles but also reshapes the basins of attraction, thereby altering the system's dynamical accessibility to different rhythmic states.}
    \label{fig4}
\end{figure}

Evidently, from Fig.~\ref{fig2}, the emergence of stable limit cycles is strongly correlated with the initial conditions of the system--an observation traditionally characterized by the concept of the \textit{basin of attraction}. In other words, the dynamical evolution in phase space at sufficiently large times can be delineated in terms of the domain of initial conditions. Notably, introducing periodic modulation in the damping factor influences this landscape, enabling transitions or switching between coexisting limit cycles, provided their dynamical stability is preserved.

\begin{figure}[h]
    \centering
    \includegraphics[width=0.8\linewidth]{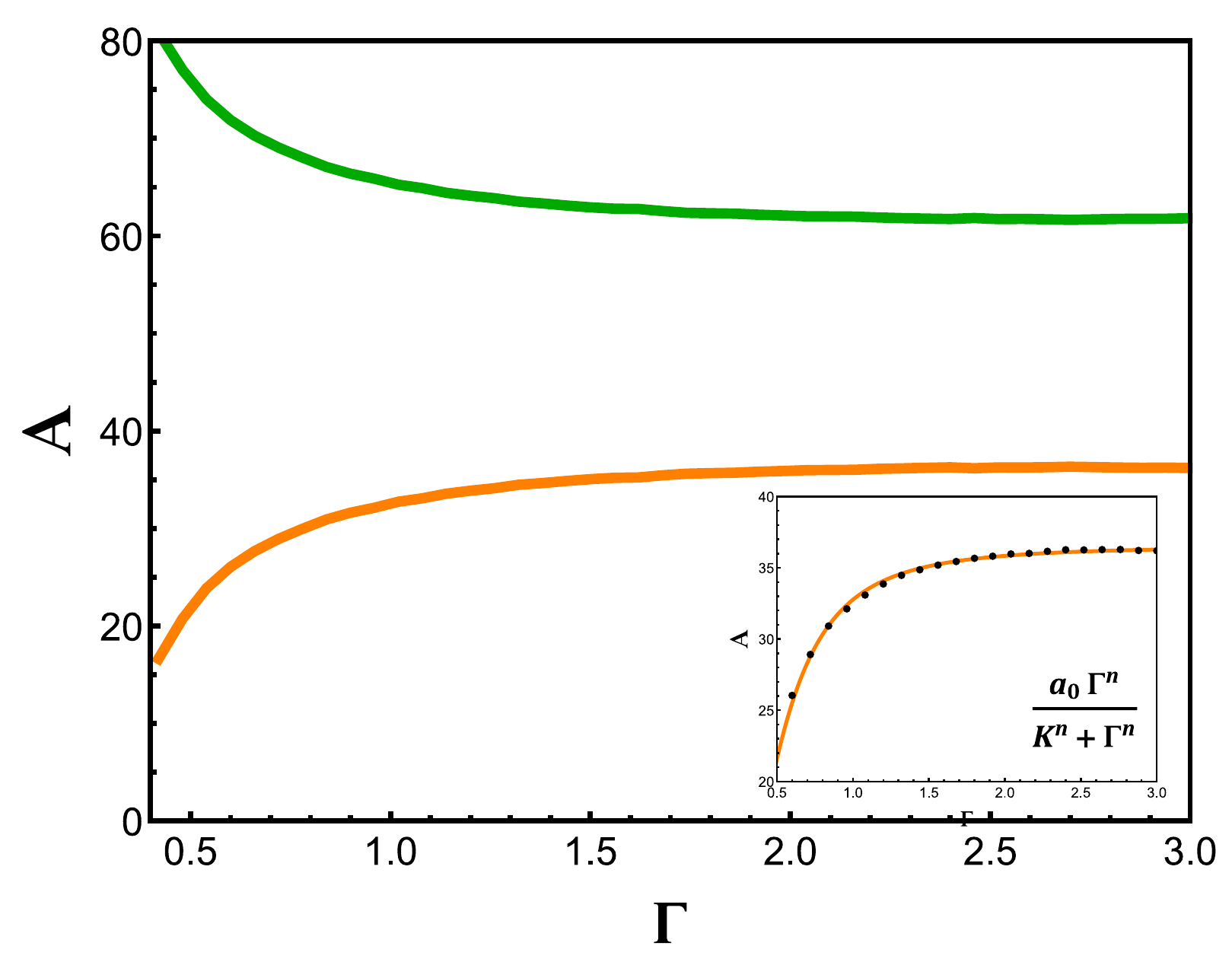}
    \caption{Variation of domain of initial condition, that corresponds to the large and intermediate limit cycle ($\Omega=6$) in terms of the modulation strength. The solid green line describes the domain of the initial condition in which the dynamics stays at intermediate amplitude. Whereas the solid orange line describes the domain of the initial condition, where the system exhibits dynamics on the large cycle. In the inset figure, we show the variation of the area fitted with the \textit{hill} function with the following set of parameters $a_0=36.521,~K=0.434~\&~n=2.5923$. The quality of this fit suggests that the growth of the intermediate-amplitude domain follows a saturating, sigmoidal profile typical of Hill-type dynamics. Notably, the two domains do not converge as $\Gamma$ increases--each remains distinct, implying the persistence of coexisting oscillatory behaviors with their respective amplitudes, sustained across a range of modulation strengths.}
    \label{fig5}
\end{figure}

To gain deeper insight into this switching behavior, we illustrate the structure of the basin of attraction as a function of the modulation strength $\Gamma$ in Fig.~\ref{fig4}(a) for $\Omega=6$, which describes the evolution of domains of initial conditions in terms of $\Gamma$. Note that $\Gamma=0$ recovers the underlying dynamics, in which we observe three distinct domains of initial conditions subjected to three stable limit cycles (see Fig.~\ref{fig4}(c)). As $\Gamma$ increases to moderate values, we observe a notable dynamical transformation: domains of initial conditions of the small and intermediate limit cycles begin to merge, indicating the collapse of their amplitudes into a unified oscillatory behavior. Interestingly, throughout this modulation-induced transition, the largest limit cycle remains robust and unaffected. This resilience highlights its dominance in the modified phase space structure.


Although the basin of attraction offers deeper insight into the underlying control mechanisms, its numerical evaluation is computationally expensive for a fixed value of $\Omega$. To circumvent this limitation, we instead examine the variation of the domain area $A$ as a function of $\Gamma$. Our analysis reveals that the region of initial conditions leading to intermediate-amplitude limit cycles initially expands with increasing $\Gamma$, eventually saturating to an asymptotic value. Interestingly, this behavior aligns well with a Hill functional form, as illustrated in the inset of Fig.~\ref{fig5}. A similar, but inverse, trend is observed for the domain corresponding to large-amplitude limit cycles: it decreases gradually with $\Gamma$ before stabilizing at an asymptotic limit.


\subsection{Justification through van der Pol plane}
\begin{figure*}[ht]
    \centering
    \includegraphics[width=\linewidth]{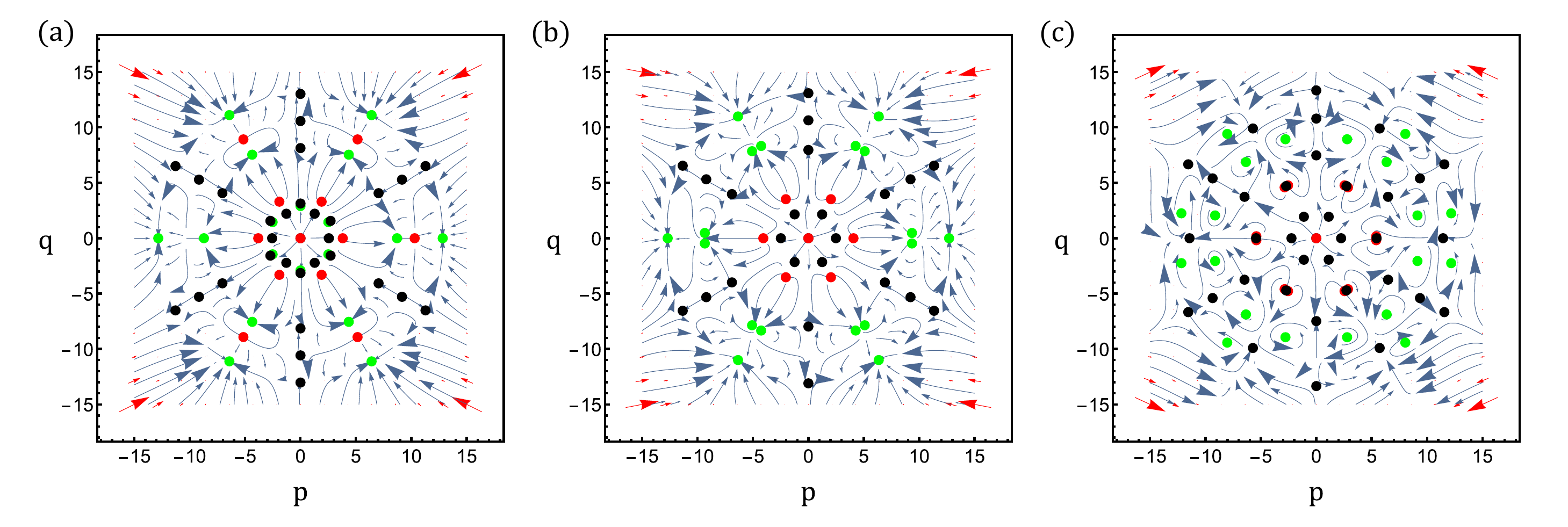}
    \caption{Representation of the stability of the limit cycles on the inferred \( (p,~q) \) plane in terms for different modulation strength for $\Omega=6$. (a) is plotted with $\Gamma=0.2$, (b) describes $\Gamma=0.4$ and (c) stands for $\Gamma=1.6$. Green points are stable limit cycles, reds are unstable and the black points are saddle points. A careful observation reveals three different class of limit cycle with distinct values of $\sqrt{p^2+q^2}$ in (a). Hence, this concludes tri-thyrhmicity. Whereas in (b) and (c), we observe only two classes of stable limit cycles, effectively this signifies bi-rhythmicity. Previously in Fig.~\ref{fig5}, we presented the domains of existence of limit cycles with different amplitudes by varying $\Gamma$.}
    \label{fig6}
\end{figure*}

The amplitude variations of the limit cycles can be analyzed within the framework of the \( (p,~q) \) phase-space. One can define the new coordinates as $p=\overline{r}\cos \overline{\phi}$ and $q=\overline{r}\sin \overline{\phi}$,      inversely $(\overline{r},~\overline{\phi})=(\sqrt{p^2+q^2},~\tan^{-1}(q/p))$. Substituting these relations in Eqs.~\eqref{dyn-avg-amp} \text{and} \eqref{dyn-avg-phi}, one gets a series of dynamics in $(p,~q)$-plane for \textit{six} different resonating frequencies. For brevity, we provide those equations in Appendix~\ref{sm-vdp}. We begin with the $x-\dot{x}$ phase plane and, to analyze the amplitude dynamics, we transition to polar coordinates using a perturbative approach. To further investigate the rhythmic behavior of the system, we adopt another coordinate framework known as the van der Pol plane. Notably, there exists a meaningful correspondence between the van der Pol plane and the original coordinate system: a stable fixed point in the van der Pol plane corresponds to a stable limit cycle in the original system. This relationship holds true for a specific class of fixed points.

For this report, we are just picking only one case for instance $\Omega=6$. Conveniently, the origin $(p,q)=(0,0)$ remains an unstable fixed point for all resonating cycles. On the other hand, performing stability analysis, we get three distinct points on the \( (p,~q) \) plane, \textit{i.e.}, (i) stable node (green dots), (ii) unstable node (red dots), and (iii) saddle node (black dots). Notably in Fig.~\ref{fig6}, three distinct classes of stable fixed points are identified in the van der Pol plane, implying the presence of trirhythmicity in the original system. As the strength of modulation increases, one class of stable fixed points disappears, leading to birhythmicity. This demonstrates that multirhythmicity in the original system can be controlled or tuned via modulation, but the system cannot exhibit monorhythmicity. \textit{Essentially, this indicates that one requires a step-by-step approach to achieve lower-order rhythmicity and does not get explosive transition towards monorhythmic nature.}

\section{Summary and discussions}\label{conclusion}
In summary, we have investigated the effect of the periodic modulation on the vdP oscillator. Compared to the underlying dynamics, we have demonstrated the branching of the stable limit cycle into two concentric limit cycles at two distinct frequencies. In particular, the system exhibits oscillation with smaller and larger amplitudes, compared to the underlying dynamics; traditionally known as anti-resonance and resonance, respectively. In the continuation, we also provide the behavior of these amplitudes with the strength of the modulation. Also, we have described how dynamical behavior changes with the delay-induced mechanism. In this context, another branching of the limit cycles has been observed for the anti-resonance branch. However, we found the dramatic change in the amplitude of the limit-cycle with the larger frequency that corresponds to the suppressed oscillation. Although, delay structures the amplitude, but it becomes less prominent due to increment of spurious degrees of freedom in the system. 

Moreover, effect of periodic modulation in the system becomes more prominent on our prototype model. To demonstrate the dynamics of the average amplitude, we invoke Krylov-Bogoliubov formalism where we observe the switching of dynamics among regions of three stable limit cycles in the presence of periodic modulation. Apriori these three concentric limit cycles was described through the framework of effective potential of the underlying dynamics. Following this route, we differentiate the domains of initial conditions in terms of $\Gamma$, where such switching can be observed. The merging phenomena of the domains has been examined, which follows the same behavior as hill function. In this scenario, we have provided a detailed numerical analysis. To support the numerical observations, we have used an analytical approach by introducing another plane, referred to as the \textit{van der Pol plane} to perform the stability analysis. In this framework, there exists a mapping between the original Cartesian coordinates and those in the van der Pol plane. A stable fixed point---or a class of stable fixed points---in the van der Pol plane corresponds to a stable limit cycle in the original \( x\text{-}\dot{x} \) plane. Similarly, a stable limit cycle---or a class of such cycles---in the van der Pol plane would manifest as oscillatory behavior of limit cycles in the \( x\text{-}\dot{x} \) plane. In this report, we focus only on the stable fixed points in the \( (p,~q) \) plane, as no periodic behavior has been observed in the van der Pol plane.

In the past, empirical observations have played a significant role in the theoretical modeling of chemical reactions, leading to the development of mathematical frameworks that forecast and explain experimental results.  Continuing this tradition, the current study offers a fresh viewpoint by emphasizing active control of reaction dynamics rather than just prediction.  We specifically investigate ways to control the temporal evolution of these systems in order to produce the desired dynamical behaviors.

Because of internal feedback loops or external environmental impacts, many chemical and biological systems naturally display periodic or quasi-periodic modulation.  In light of this, our research explores the effects of implementing controlled internal modulation as a strategy for managing multirhythmic behavior in systematic or step-by-step transitions towards lower rhythmicity.  We contend that this method provides significant understanding of the composition and changes of intricate dynamical states.  Our findings are important because they may be applied to a wide range of systems in which nonlinear oscillatory dynamics is important.  Our paradigm provides a potential path for future experimental and theoretical investigation in the management of complex response dynamics by utilizing inherent modulation processes to open up new channels for focused regulation of system evolution.

\bibliography{reference.bib}

\onecolumngrid
\appendix
\section{Average amplitude and phase dynamics}\label{appendixA}

In this section, we present the mathematical expressions of Eqs.~\eqref{dyn-avg-amp} and \eqref{dyn-avg-phi}, corresponding to the parametric modulation for various modulation frequencies.

\begin{align}
    \left.\mathcal{A}\right|_{\Omega=2}&=-\mu\dfrac{\Gamma}{512}\overline{r}(128-8\alpha \overline{r}^4+8\beta \overline{r}^6-7\gamma \overline{r}^8+6\delta\overline{r}^{10})\cos[2\overline{\phi}],\\
    \left.\mathcal{B}\right|_{\Omega=2}&=\mu\dfrac{\Gamma}{1024}(256-128\overline{r}^2+80\alpha \overline{r}^4-56\beta \overline{r}^6+42\gamma \overline{r}^8-33\delta \overline{r}^{10})\sin[2\overline{\phi}].\\
    \left.\mathcal{A}\right|_{\Omega=4}&=\mu\dfrac{\Gamma}{4096}\overline{r}^3(256-128\alpha \overline{r}^2+64\beta \overline{r}^4-32\gamma \overline{r}^6+15\delta \overline{r}^8)\cos[4\overline{\phi}],\\
    \left.\mathcal{B}\right|_{\Omega=4}&=-\mu\dfrac{\Gamma}{4096}\overline{r}^2(256-256\alpha\overline{r}^2+224\beta \overline{r}^4-192\gamma\overline{r}^6+165\delta\overline{r}^8)\sin[4\overline{\phi}].\\
    \left.\mathcal{A}\right|_{\Omega=6}&=-\mu\dfrac{\Gamma}{1024}\overline{r}^5(16\alpha-16\beta \overline{r}^2+13\gamma\overline{r}^4-10\delta\overline{r}^{6})\cos[6\overline{\phi}],\\
    \left.\mathcal{B}\right|_{\Omega=6}&=\mu\dfrac{\Gamma}{2048}\overline{r}^4(32\alpha-48\beta\overline{r}^2+54\gamma\overline{r}^4-55\delta \overline{r}^6)\sin[6\overline{\phi}].\\
    \left.\mathcal{A}\right|_{\Omega=8}&=\mu\dfrac{\Gamma}{2048}\overline{r}^7(8\beta-12\gamma \overline{r}^2+13\delta \overline{r}^4)\cos[8\overline{\phi}],\\
    \left.\mathcal{B}\right|_{\Omega=8}&=-\mu\dfrac{\Gamma}{1024}\overline{r}^6(4\beta-8\gamma \overline{r}^2+11\delta \overline{r}^4)\sin[8\overline{\phi}].\\
    \left.\mathcal{A}\right|_{\Omega=10}&=-\mu\dfrac{\Gamma}{1024}\overline{r}^9(\gamma-2\delta\overline{r}^2)\cos[10\overline{\phi}],\\
    \left.\mathcal{B}\right|_{\Omega=10}&=\mu\dfrac{\Gamma}{2048}\overline{r}^8(2\gamma-5\delta \overline{r}^2)\sin[10\overline{\phi}].\\
    \left.\mathcal{A}\right|_{\Omega=12}&=\mu\dfrac{\Gamma}{4096}\delta \overline{r}^{11}\cos[12\overline{\phi}],\\
    \left.\mathcal{B}\right|_{\Omega=12}&=-\mu\dfrac{\Gamma}{4096}\delta\overline{r}^{10}\sin[12\overline{\phi}].
\end{align}

\section{Dynamics on the Van der Pol plane}\label{sm-vdp}

In this section, we provide the analytical expressions of the dynamics on the Van der Pol plane for various modulation frequencies. 

\begin{align}
    \left.\dot{p}\right|_{\Omega=2}&=\dfrac{\mu p}{1024}\left(-256(\Gamma-2)-3(4\Gamma+7)\delta p^{10}+2p^4(8\alpha(\Gamma+4)+12\beta(4\Gamma-5)q^2+42\gamma(2-3\Gamma)q^4+105(2\Gamma-1)\delta q^6)\right.\nonumber\\
    &+p^2(15(20\Gamma-7)\delta q^8-8(4\alpha(5\Gamma-4)q^2+15\beta(1-2\Gamma)q^4+7\gamma(5\Gamma-2)q^6+16))\nonumber\\
    &+2p^6(-4\beta(2\Gamma+5)-28\gamma(\Gamma-2)q^2+15(8\Gamma-7)\delta q^4)+p^8(14\gamma(\Gamma+2)+15(2\Gamma-7)\delta q^2)\nonumber\\
    &+\left.16\alpha(4-11\Gamma)q^4+8\beta(16\Gamma-5)q^6+14\gamma(2-7\Gamma)q^8+3(26\Gamma-7)\delta q^{10}+128(2\Gamma-1)q^2\right),\\
    \left.\dot{q}\right|_{\Omega=2}&=-\dfrac{\mu q}{1024}\left(-256(\Gamma+2)3(26\Gamma+7)\delta p^{10}+2p^4(-8\alpha(11\Gamma+4)+60q^2(2\beta+\Gamma+\beta)-42\gamma(3\Gamma+4)q^4+15(8\Gamma+7)\delta q^6)\right.\nonumber\\
    &+p^2(128(2\Gamma+1)-32\alpha(5\Gamma+4)q^2+24\beta(4\Gamma+5)q^4-56\gamma (\Gamma+2)q^6+15(2\Gamma+7)\delta q^8)\nonumber\\
    &+2p^6(4\beta(16\Gamma+5)-28\gamma(5\Gamma+2)q^2+105(2\Gamma+1)\delta q^4)+p^8(15(20\Gamma+7)\delta q^2-14\gamma(7\Gamma+2))\nonumber\\
    &\left.+2q^2(8\alpha(\Gamma-4)q^2+4\beta(5-2\Gamma)q^4+7\gamma(\Gamma-2)q^6+64)+3(7-4\Gamma)\delta q^{10}\right).\\
    \left.\dot{p}\right|_{\Omega=4}=&\dfrac{\mu p}{4096}\left(3(5\Gamma-28)\delta p^{10}+2p^4(-16(4\alpha(\Gamma-2)+3\beta(5-6\Gamma)q^2+7\gamma(2\Gamma-3)q^4)-105(\Gamma+4)\delta q^6)\right.\nonumber\\
    &+p^2(32(8(\gamma-2)-8\alpha(\Gamma-2)q^2-5\beta(2\Gamma+3)q^4)+14q^6(2\gamma\Gamma+\gamma)-105(13\Gamma+4)\delta q^8)+\nonumber\\
    &+2p^6(16\beta(2\Gamma-5)=32\gamma(7-10\Gamma)q^2+15(37\Gamma-28)\delta q^4)+p^8(16\gamma(7-2\Gamma)+15(41)\Gamma-28)\delta q^2)\nonumber\\
    &+\left.16(8\alpha(7\Gamma+2)q^4-2\beta(26\Gamma+5)q^6+\gamma(46\Gamma+7)q^8-16(3\Gamma+2)q^2+128)-3(215\Gamma+28)\delta q^{10}\right),\\
    \left.\dot{q}\right|_{\Omega=4}=&-\dfrac{\mu q}{4096}\left(3(215\Gamma+28)\delta p^{10}+2p^4(16(-4\alpha(7\Gamma+2)+5\beta(2\Gamma+3)q^2+7\gamma(2\Gamma-3)q^4)+15(28-37\Gamma)\delta q^6\right.\nonumber\\
    &+p^2(32(8(3\Gamma+2)+8\alpha(\Gamma-2)q^2+3\beta(5-6\Gamma)q^4+2\gamma(10\Gamma-7)q^6)+15(28-41\Gamma)\delta q^8)\nonumber\\
    &+\left.16(8\alpha(\Gamma-2)q^42\beta(5-2\Gamma)q^6+\gamma(2\Gamma-7)q^8-16(\Gamma-2)q^2-128)+3(28-5\Gamma)\delta q^{10}\right).\\
    \left.\dot{p}\right|_{\Omega=6}=&\dfrac{\mu p}{1024}\left((10\Gamma-21)\delta p^{10}-2p^4(8\alpha(\Gamma-4)+3q^2(4\beta(4\Gamma+5)-7\gamma(9\Gamma+4)q^2+35(3\Gamma+1)\delta q^4))\right.\nonumber\\
    &+p^2(4(8\alpha(5\Gamma+4)q^2-30q^4(2\beta\Gamma+\beta)+7\gamma(7\Gamma+4)q^6-32)-15(6\Gamma+7)\delta q^8)\nonumber\\
    &+ 2p^6(4\beta(2\Gamma-5)+2\gamma(5\Gamma+28)q^2-15(12\Gamma+7)\delta q^4)+p^8(\gamma(28-13\Gamma)+35(\Gamma-3)\delta q^2)+\nonumber\\
    &+16\alpha(4-5\Gamma)q^4+8\beta(16\Gamma-5)q^ 6+\gamma(28-149\Gamma)q^8+(155\Gamma-21)\delta q^{10}-128q^2+512 \big),\\
    \left.\dot{q}\right|_{\Omega=6}=&-\dfrac{\mu q}{1024}\big((155\Gamma+21)\delta p^{10}-2p^4(8\alpha(5\Gamma+4)+60\beta(2\Gamma-1)q^2+21\gamma(4-9\Gamma)q^4+15(12\Gamma-7)\delta q^6)\nonumber\\
    &+p^2(4(8\alpha(5\Gamma-4)q^2+6\beta(5-4\Gamma)q^4+\gamma(5\Gamma-28)q^6+32)+35(\Gamma+3)\delta q^8)+\nonumber\\
    &+2p^6(4\beta(16\Gamma+5)+14\gamma(7\Gamma-4)q^2+105(1-3\Gamma)\delta q^4)+p^8(15(7-6\Gamma)\delta q^2-\gamma(149\Gamma+28))\nonumber\\
    &-16\alpha(\Gamma+4)q^4+8\beta(2\Gamma+5)q^6-\gamma(13\Gamma+28)q^8+(10\Gamma+21)\delta q^{10}+128q^2-512\big).\\
    \left.\dot{p}\right|_{\Omega=8}=&\dfrac{\mu p}{2048}\big((13\Gamma-42)\delta p^{10}+2p^4(64\alpha-12\beta(7\Gamma+10)q^2+28\gamma(\Gamma+6)q^4+21(13\Gamma-10)\delta q^6)\nonumber\\
    &+p^2(256\alpha q^2+40\beta(7\Gamma-6)q^4+112\gamma(2-5\Gamma)q^6+15(47\Gamma-14)\delta q^8-256)\nonumber\\
    &+2p^6(4\beta(\Gamma-10)+8\gamma(13\Gamma+14)q^2-15(17\Gamma+14)\delta q^4)+p^8(\gamma(56-12\Gamma)-35(5\Gamma+6)\delta q^2)\nonumber\\
    &+4(32\alpha q^4-2\beta(7\Gamma+10)q^6+\gamma(29\Gamma+14)q^8-64q^2+256)-(163\Gamma+42)\delta q^{10}\big),\\
    \left.\dot{q}\right|_{\Omega=8}=&\dfrac{\mu q}{2048}\big(-(163\Gamma+42)\delta p^{10}+2p^4(64\alpha+20\beta(7\Gamma-6)q^2+28\gamma(\Gamma+6)q^4-15(17\Gamma+14)\delta q^6)\nonumber\\
    &+p^2(8(32\alpha q^2-3\beta(7\Gamma+10)q^4+2\gamma(13\Gamma+14)q^6-32)-35(5\Gamma+6)\delta q^8)\nonumber\\
    &+2p^6(-4\beta(7\Gamma+10)+56\gamma(2-5\Gamma)q^2+21(13\Gamma-10)\delta q^4)+p^8(4\gamma(29\Gamma+14)+15(47\Gamma-14)\delta q^2)\nonumber\\
    &+4(32\alpha q^4+2\beta(\Gamma-10)q^6+\gamma(14-3\Gamma)q^8-64q^2+256)+(13\Gamma-42)\delta q^{10}\big).
\end{align}
\begin{align}
    \left.\dot{p}\right|_{\Omega=10}=&\dfrac{\mu p}{1024}\big((2\Gamma-21)\delta p^{10}+2p^4(32\alpha-60\beta q^2+21\gamma(4-3\Gamma)q^4+105(\Gamma-1)\delta q^6)\nonumber\\
    &+p^2(128\alpha q^2-120\beta q^4+28\gamma(3\Gamma+4)q^6-105q^8 (2\Gamma+\delta+\delta)-128)+p^6(-40\beta+4\gamma(9\Gamma+28)q^2+30(4\Gamma-7)\delta q^4)\nonumber\\
    &-p^8(\gamma(\Gamma-28)+5(13\Gamma+21)\delta q^2)+64alpha q^4-40\beta q^6+\gamma(28-9\Gamma)q^8+(23\Gamma-21)\delta q^{10}-128q^2+512\big),\\
    \left.\dot{q}\right|_{\Omega=10}=&-\dfrac{\mu q}{1024}\big((23\Gamma+21)\delta p^{10}+2p^4(-32\alpha+60\beta q^2-21\gamma(3\Gamma+4)q^4+15(4\Gamma+7)\delta q^6)\nonumber\\
    &+p^2(4(-32\alpha q^2+30\beta q^4+\gamma(9\Gamma-28)q^6+32)+5(21-13\Gamma)\delta q^8)+2p^6(20\beta+14\gamma(3\Gamma-4)q^2+105(\Gamma+1)\delta q^4)\nonumber\\
    &+p^8(105(1-2\Gamma)\delta q^2-\gamma(9\Gamma+28))-64\alpha q^4+40\beta q^6-\gamma(\Gamma+28)q^8+(2\Gamma+21)\delta q^{10}+128 q^2-512\big).\\
    \left.\dot{p}\right|_{\Omega=12}=&\dfrac{\mu p}{4096}\big(32(8\alpha(p^2+q^2)^2-5\beta(p^2+q^2)^3-16(p^2+q^2-4))\nonumber\\
    &+112\gamma(p^2+q^2)^4+\delta(\Gamma(-55p^8q^2+330p^4q^6+165p^2q^8+p^{10}-11q^{10})-84(p^2+q^2)^5)\big),\\
    \left.\dot{q}\right|_{\Omega=12}=&\dfrac{\mu q}{4096}\big(32(8\alpha(p^2+q^2)^2-5\beta(p^2+q^2)^3-16(p^2+q^2-4))\nonumber\\
    &+112\gamma(p^2+q^2)^4-\delta(\Gamma(-165p^8q^2+462p^6q^4-330p^4q^6+55p^2q^8+11p^{10}-q^{10})+84(p^2+q^2)^5)\big).
\end{align}
\end{document}